# Si plate radius influence on the photoacoustic signal processed by neural networks

Katarina Lj Djordjevic[1], Dragana K Markushev[2], Slobodanka P Galović[1], Dragan D Markushev[2], Jose Ordonez-Miranda[3]

[1]University of Belgrade, „VINČA" Institute of Nuclear Sciences - National Institute of the Republic of Serbia, University of Belgrade, PO box 522, 11000 Belgrade, Serbia
[2]University of Belgrade, Institute of Physics Belgrade, National Institute of the Republic of Serbia, Pregrevica 118, 11080 Belgrade (Zemun), Serbia
[3]LIMMS, CNRS-IIS UMI 2820, The University of Tokyo, Tokyo 153-8505, Japan.

Katarina Djordjevic e-mail: katarina.djordjevic@vin.bg.ac.rs

**Abstract**

The effect of the sample radius on the total photoacoustic signal processed by neural networks trained with undistorded and distorded signals is carefully analized for modulation frequencies from 20 to 20 kHz. This is done for signals generated for a 400-µm-thick Si n-type plate, whose radius varies from 2 to 7 mm. It is found that the networks trained with both undistorded or distorded signals yield the best predictions for sample radii between 2 and 3 mm, which is close to the used microphone aperture radius of 1.5 mm. The network trained only with undistorded signals gives the best results for sample radii comparable to the microphone dimensions. The obtained results justify the validity of a theoretical model derived for a sample radius equals to the microphone aperture and indicate the experimental necessity to use samples with radii comparable to this aperture.

## 1. Introduction

The photoacoustic effect is the effect of the sound appearance due to the illumination of a sample with an optical beam whose intensity is time-dependent [1-5]. Numerous non-destructive methods for characterization and imaging of various solids, multilayer structures, microelectronic and photonic devices, soft matter and biological tissues are based on this effect [6-21]. The use of neural networks in photoacoustics has become very popular recently, especially imaging witin biomedical application [23-25]. Significant progress has been made, also, in photoacoustics of solids with different types of networks both in material characterization and in the development of experimental procedures [26-28]. Networks have proven to be a very effective tool for predicting the physical characteristics of solid samples [29-31]. Within the framework of multi-parameter fitting in photoacoustics, the networks most often predicted the values of the sample parameters such as: thermal diffusion ($D_T$) and thermal expansion ($\alpha_T$) together with the thickness of the sample ($l$) [32,33]. These parameters are important because they occur when calculating the temperature distributions of the sample and all components of the photoacoustic signal generated by the periodic illumination of the sample.

It has long been known that the geometric parameters of the sample are very important for the correct interpretation of the photoacoustic signal in the frequency domain [34-36]. If the circular plate shape samples are tested, one of such parameters is its radius $R_s$. It explicitly appears in the equation for calculating the thermoelastic and plasmaelastic components of the photoacoustic signal of a semiconductor, so that its influence on the shape of the total signal through these two components can be extremely significant [32,33]. In this paper, the analysis of the influence of the semiconductor sample radius changes on all components of the photoacoustic signal, thermodiffusion (TD), thermoelastic (TE) and plasmaelastic (PE), is presented, based on the theoretical model of a composite piston [1-3] that serves to form an appropriate base of photoacoustic signals for neural network training [26,27]. The tested sample is Si n-type 400 μm thick wafer, classified as a plasma-thick sample - sample with a thickness larger than the diffusion length of free carriers [32,33]. Plasma-thick semiconductor samples have a thermal response of known and characteristic patterns of behaviour of the total signal and its components in the modulation frequencies range from 20 Hz to 20 kHz [33]. These patterns include domination of TD component at lower, and TE component at higher frequencies, while PE component influence on total signal is negligible.

The basic idea of this paper is to determine which type of neural network gives better predictions in the case of a change in the radius of the sample: one that is trained by distorted signals, or by signals undistorted in relation to instrumental deviations [37-39]. This analysis is very important because its results can give the possibility of working with distorted signals, which significantly simplifies and speeds up the process of characterization of the tested samples. Both types of network training signals were obtained theoretically, and the final verification of network efficiency was tested on experimental data. The conclusions of this analysis indicate, in experimental terms, the importance of the choice of sample radius in relation to the dimensions of the microphone as a photoacoustic cell and signal detector. Also, based on the obtained networks predictions a conclusion can be drawn on the justification of using approximations in the theoretical model of the composite piston, especially those related to the equality of microphone aperture and sample radii.

## 2. Theoretical introduction

The theoretical model of the composite piston presented in this paper is adapted to the experimental setup of an open photoacoustic cell [32,33,37,40]. It is assumed that frequency-modulated light source of wavelength 660nm illuminates the Si n-type plate. Due to the photoacoustic effect of the absorbed modulated light and the sample interaction, a sound is recorded by an open photoacoustic cell in the transmission configuration (Figure 1). The photoacoustic signal detector is assumed to be an electret microphone (ECM 30B, JinInElectronicCo., Ltd.) with a radius of $R_0 = 4.9$ mm, and aperture radius of $r_0 = 1.5$ mm [33]. The microphone at the same time can serve as a measuring cell of minimal volume $V_0 = \pi r_0^2 l_c$, where $l_c$ is the length from the microphone aperture to its membrane (see Figure 1).

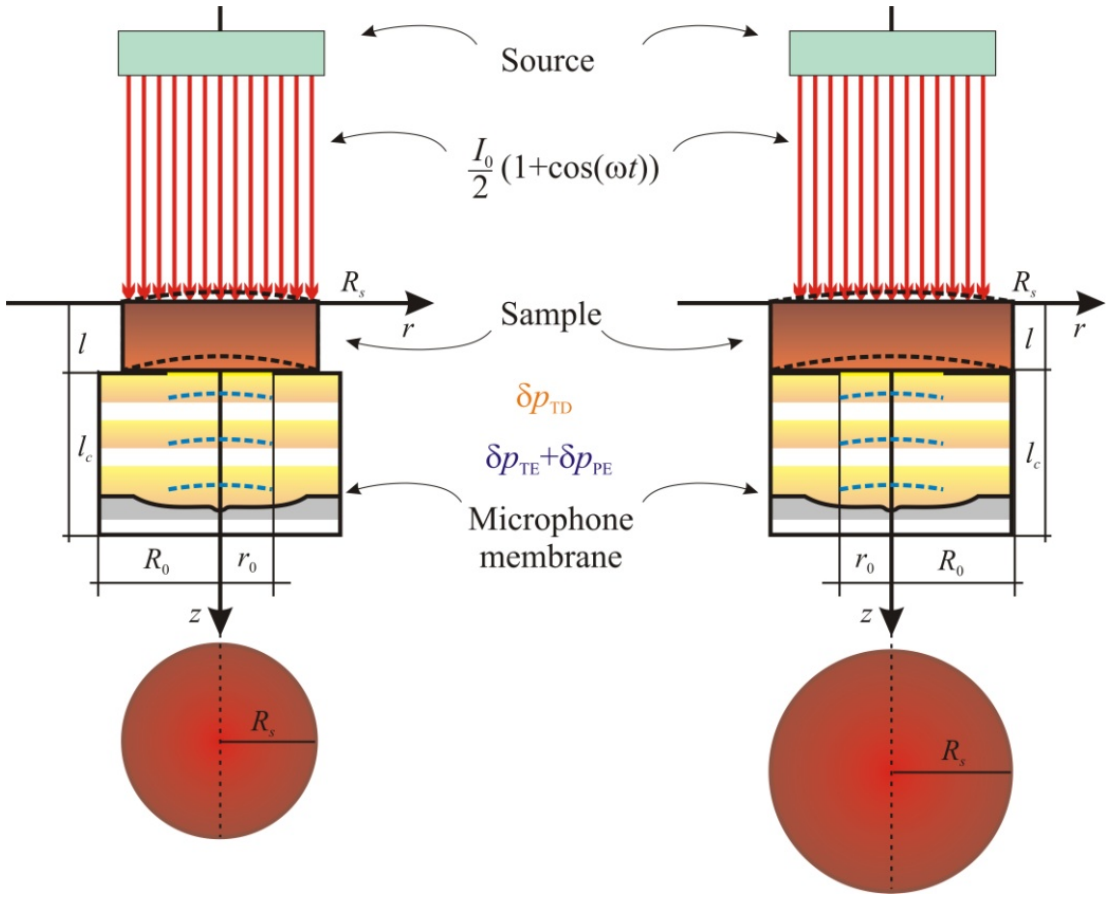

Figure 1. Simple ideal scheme of the transmission configuration of an open photoacoustic cell with an electret microphone and samples of different radii.

The measured signal in the transmission configuration and in the frequency range from 20Hz to 20 kHz is a sound signal that occurs due to the photoacoustic effect in the sample itself and is usually distorted by the influence of used instruments and the different types of noise. The signal thus measured is called the distorted signal $S_{total}(f)$. If a well-known signal correction procedure [33,37-39] is applied to such a signal, so-called `true' or undistorted signal $\delta p_{total}(f)$ can be obtained, generated from an illuminated sample. The theoretical model that fully describes the undistorted signals is based on the composite piston model [1-3]. This model explains that the total photoacoustics signal $\delta p_{total}(f)$ of a semiconductor is equal to the sum of its components; thermodiffusion $\delta p_{TD}(f)$, thermoelastic $\delta p_{TE}(f)$ and plasmaelastic $\delta p_{PE}(f)$ [33,40]:

$$\delta p_{total}(f) = \delta p_{TD}(f) + \delta p_{TE}(f) + \delta p_{PE}(f). \tag{1}$$

The thermodiffusion component can be written as follows [33,41]:

$$\delta p_{TD}(t) = \frac{\gamma p_0}{l_c} \frac{\mu_g}{\sqrt{2}T_0} T(l) e^{i\left(\omega t - \frac{\pi}{4}\right)}, \tag{2}$$

where $\omega = 2\pi f$, $f$ is modulation frequency, $\gamma$ is an adiabatic ratio, $p_0$ and $T_0$ are the ambient pressure and temperature, $\mu_g$ is the thermal diffusion length of the surrounding gas, $T_s(l)$ (see Appendix I) is the dynamic temperature component at the back sample surface, $l$ is the sample thickness and $l_c$ is the length of the cell (microphone).

Due to different temperature values on the illuminated and non-illuminated side of the silicon sample, its periodic bending and expansion occurs. The effect of bending and spreading is most simply described by cylindrical ($z$, $r$) coordinates so that the thermoelastic component $\delta p_{TE}(f)$ of the photoacoustic signal can be written in the form [33,40]:

$$\delta p_{TE}(f) = \alpha_T \frac{3\pi p_0 \gamma}{V_0} \frac{R_s^4}{l^3} \int_{-l/2}^{l/2} zT(z)dz \tag{3}$$

where $T(z)$ is the temperature distribution in the sample along the $z$-axes (see Appendix I), $\alpha_T$ is the coefficient of thermal expansion, $V_0$ is the microphone (cell) volume, and $R_s$ is the sample radius. In most of the theoretical analysis, the $R_s = r_0$ approximation is taken to simplify analysis.

By modulated illumination of the sample with the light of proper wavelength, free carriers (electrons and holes) are generated, which leads to additional stress and periodic bending of the sample. The influence of phototgenerated carriers on the bending of the sample is described by the plasmaelastic component $\delta p_{PE}(f)$ [33,40]:

$$\delta p_{PE}(f) = d_n \frac{p_0 \gamma}{V_0} \frac{3\pi R_s^4}{l^3} \int_{-l/2}^{l/2} z \delta n_p(z) dz, \quad (4)$$

where $\delta n_p(z)$ is the minority carrier density (see Appendix II), and $d_n$ is the coefficient of electronic deformation. Analysing Eq.(3&4) one can see that $\delta p_{TE}(f)$ and $\delta p_{PE}(f)$ depends on $R_s$ as $R_s^4$, meaning that small changes in $R_s$ leads to the large changes in $\delta p_{TE}(f)$ and $\delta p_{PE}(f)$

It must be bear in mind that during the measurement there are a lot of signal distortions due to the influence of the used instruments (microphones and accompanying electronics). These distortions have signal filtering characteristics and are described, in the low-frequency range, by the function of the first-order low-pass filter $H_e(f)$ in the form [37,39]:

$$H_e(f) = -\frac{\omega \tau_{c1}}{(1 + i\omega \cdot \tau_{c1})} \cdot \frac{\omega \tau_{c2}}{(1 + i\omega \cdot \tau_{c2})}, \quad (5)$$

where $\omega = 2\pi f$, $f$ is the modulation frequency, $\tau_{c1} = (2\pi f_{c1})^{-1} = R_1 C_1$ and $\tau_{c2} = (2\pi f_{c2})^{-1} = R_2 C_2$ are the time constants and $RC$ characteristics of the microphone and accompanying electronics (usually sound-card or lock-in).

In the high-frequency range, these distortions are described by the function of the second-order low-pass filter $H_a(f)$ in the form [37,39]:

$$H_a(f) = \frac{\omega_{c3}^2}{\omega_{c3}^2 + i\delta_{c3}\omega_{c3}\omega - \omega^2} + \frac{\omega_{c4}^2}{\omega_{c4}^2 + i\delta_{c4}\omega_{c4}\omega - \omega^2}, \quad (6)$$

where $\delta_k$ is the damping factor ($k = c3, c4$), $\omega_{c3}$ is the microphone cut-off frequency and $\omega_{c4}$ is the characteristic frequency which depends on the geometry of the microphone body.

The total measured signal $S_{total}(f)$ can be written as a product of $\delta p_{total}(f)$, $H_e(f)$ i $H_a(f)$ [33,37,39]:

$$S_{total}(f) = \delta p_{total}(f) H_e(f) H_a(f). \quad (7)$$

Characteristic parameters $H_e(f)$ and $H_a(f)$ can be obtained either from the microphone manual or by independent measurements [33,37,39]. By finding them, $H_e(f)$ and $H_a(f)$ can be removed from $S_{total}(f)$ in Eq. (7) so the required $\delta p_{total}(f)$ can be obtained [33,37], which is further described by the composite piston theory (Eqs.(1-4)).

## 3. Neural networks

Our motive for the study of neural networks is their use in the precise characterization of the properties of the tested materials. In the process of supervised learning by connecting the data of input neurons (values of photoacoustic signals), with the data of output neurons (radius), the weights are adjusted to the moment when the output becomes increasingly like the desired data values. In our work with networks so far, we have learned that for the sake of prediction, we need to present undistorted data to the trained network, data that are adapted to the theoretical model used to create the training base. This assumes that during the processing of experimental results, we must also count on the signal correction procedure. If it turns out that signal correction is not required to determine some of the sample parameters (expected in the case of only $R_s$ changes), this will mean significant time savings. Therefore, in this paper, we will try to train networks with distorted (NN1) and undistorted (NN2) photoacoustic signals with the desire to determine the radius of the investigated sample, as a very important experimental parameter. Our goal is to see if there are differences in the prediction of the radius of the sample of such formed networks and whether, based on that, we can speed up the process of processing experimental data.

For the sake of simplicity, and based on the problem we want to solve, we decided that both networks used here have the simplest architecture, the so-called shallow networks: input layer, one hidden layer, and output layer [26,27]. The input layer was determined by the 144 neurons ($i = 72$ values of the photoacoustic signal amplitude $A_i$, and $i = 72$ values of photoacoustic signal phase $\varphi_i$ characteristic), the hidden layer was determined by 10 neurons, and the output layer was determined by one neuron (the prediction of $R_s$, Figure 2). The choice of neural network architecture ensures good accuracy and optimal training time. Due to the better precision of the neural network, we adjusted the amplitudes by logarithmic normalization $20\log_{10}A_i$ to obtain the phase values range ($A_i$ represents amplitudes normalized to unit pressure).

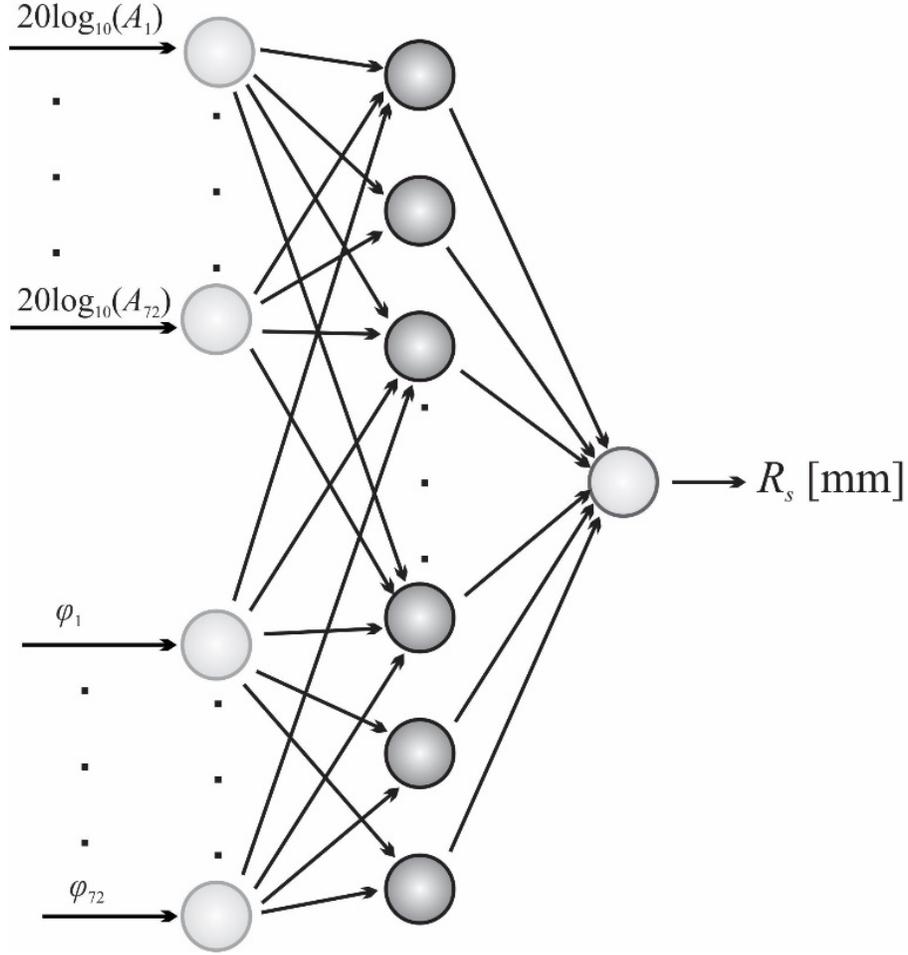

Figure 2. Representation of a single-layer neural network used for the analysis of distorted and undistorted photoacoustic signals that depend on the radius of silicon samples.

## 4. Results and discussion

### 4.1. Signal database formation

As it was stated in the previous paragraphs, analizing Eqs. (3&4) one can observe that the thermoelastic $\delta p_{TE}(f)$ and plasmaelastic $\delta p_{PE}(f)$ components of the total photoacoustic signal depend on the geometric parameter of the sample radius as $R_s^4$. Due to the high sensitivity of these components to $R_s$ changes in the total photoacoustic signal, we formed a `small` bases of 26 undistorted $\delta p_{total}(f)$ and 26 distorted $S_{total}(f)$ theoretical photoacoustic signals obtained by Eqs.(1-7). As we said earlier, we assume that the sample is Si n-type plate, 400 μm thick. Bases are determined by amplitude-phase characteristics in the frequency domain from 20 Hz to 20 kHz, changing the $R_s$ from 2 mm to 7 mm in the steps of 0.2 mm. To simplify our considerations, we assumed that the excitation light intensity $I_0$ = const. The dependences of the a) amplitudes and b) phases of the $S_{total}(f)$ and $\delta p_{total}(f)$ signals on $R_s$ in the frequency domain are clearly shown in Figures 3 and 4, respectively. The influence of the $R_s$ is clearly expressed in both Figures at frequencies $f > 100$ Hz in the amplitudes (saddle shape) and in phases (distend areas). The black arrows indicate the directions of changes due to the $R_s$ value increase. Unchanged amplitudes and phases in the low-frequency region ($f < 100$ Hz) indicate an area where the $S_{total}(f)$ and $\delta p_{total}(f)$ are insensitive to $R_s$ changes.

Figure 3 clearly shows the instrumental influences as electronic filtering (Eq. (5)) at lower ($H_e$ at $f < 100$ Hz) and acoustic filtering (Eq. (6)) at higher frequencies ($H_a$ at $f > 5$ kHz) [33,37-39]. Electronic filters $H_e$ reduce amplitudes (Figure 3.a) and raise phases (Figure 3.b) at low frequencies. Acoustic filters are characterized by resonant peaks in both amplitudes and phases. The parameters that define $H_e$ and $H_a$ are not dependent on $R_s$ and are constant during its changes. Therefore, the reasons for changing $S_{total}(f)$ and $\delta p_{total}(f)$ by changing $R_s$ are the same, and consist of the following: both signals are characterized with a seddle shape in amplitudes, which is more pronounced for $R_s = 2$ mm, while with an increasing radius the saddle shape slowly disappears and at $R_s = 7$ mm completely vanishes.

The insensitivity of the $S_{total}(f)$ and $\delta p_{total}(f)$ signals to $R_s$ changes at lower frequencies allows one the possibility of, so-called, signal self-normalization that can contribute to the better performances of network training. On the other hand, the saddle shape of the $S_{total}(f)$ and $\delta p_{total}(f)$ signals is a typical pattern of plasma-thick silicon samples behavior in frequency domain (strong TD and TE influence), allowing one to obtain better network parameter prediction. The both mentione signal characteristics, insensitivity, and saddle shape, are the reason why, in this paper, we decided to create a training base with distorted and undistorted signals of plasma-thick sample, expecting the efficient neural networks optimization.

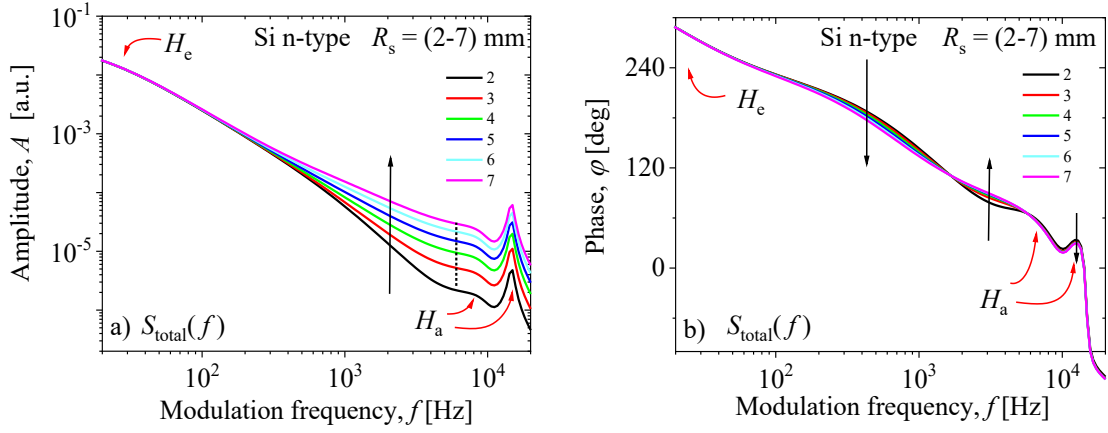

Figure 3. The bases of distorted PA signal $S_{total}(f)$: a) amplitudes, $A$, and b) phase, $\varphi$, from Si n-type as a function of simple radius $R_s$ and modulation frequency, $f$.

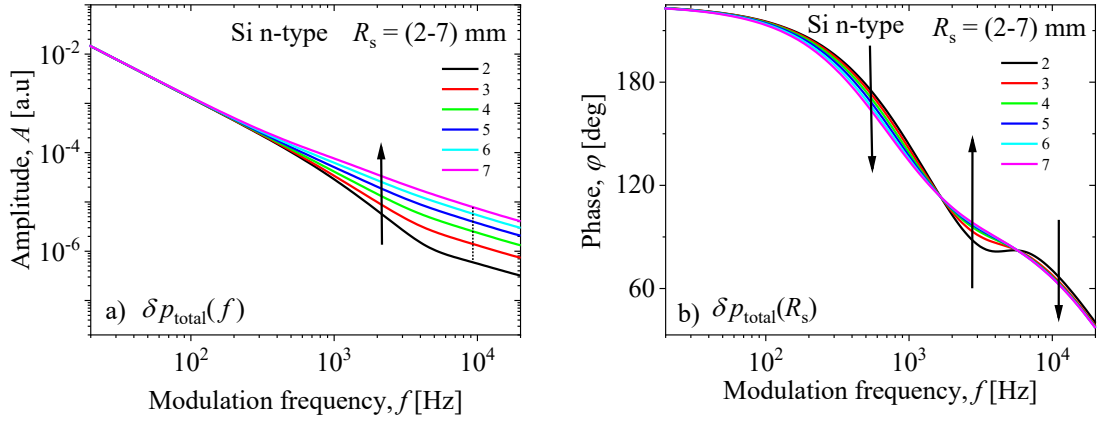

Figure 4. The bases of undistorted PA signal $\delta p_{total}(f)$: a) amplitudes, $A$, and b) phase, $\varphi$, from Si n-type as a function of simple radius $R_s$ and modulation frequency, $f$.

The question may be asked why the signals $S_{total}(f)$ and $\delta p_{total}(f)$ changes in amplitudes and phases due to $R_s$ variations are most visible at $f > 100$ Hz. The explanation can be given by the theory based on the composite piston model (Eqs. (1-4)), in which the radius of the sample does not affect all components of the photoacoustic signal equally. Changes in $R_s$ do not affect changes in the amplitude of $\delta p_{TD}(f)$ component at all (Figure 5.a), but they do change the amplitudes of $\delta p_{TE}(f)$ (Figure 6.a) and $\delta p_{PE}(f)$ components (Figure 7.a). In both $\delta p_{TE}(f)$ and $\delta p_{PE}(f)$ cases, the amplitudes increase with $R_s$. In all cases (Figures 5.b, 6.b & 7.b) phases do not change at all.

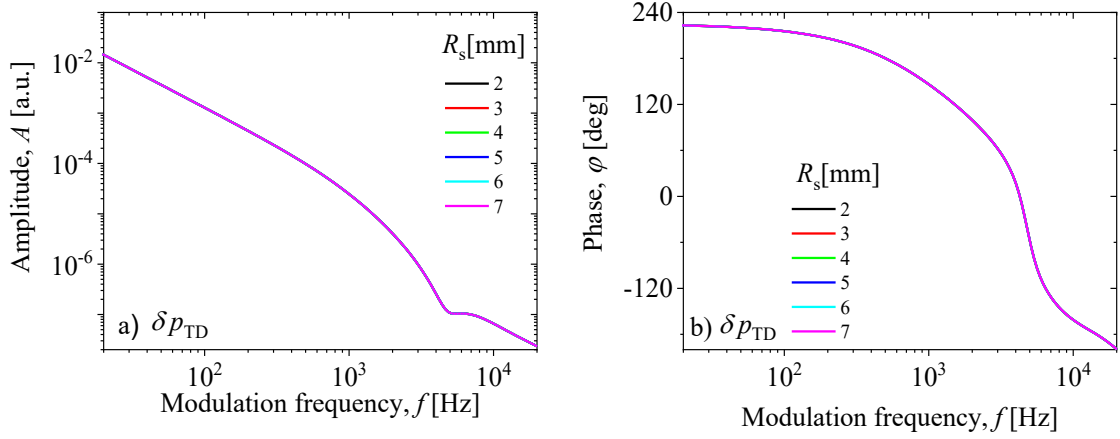

Figure 5. Thermodiffusion component $\delta p_{TD}(f)$ a) amplitudes $A$ and b) phases $\varphi$ as a function of simple radius $R_s$, and modulation frequency, $f$.

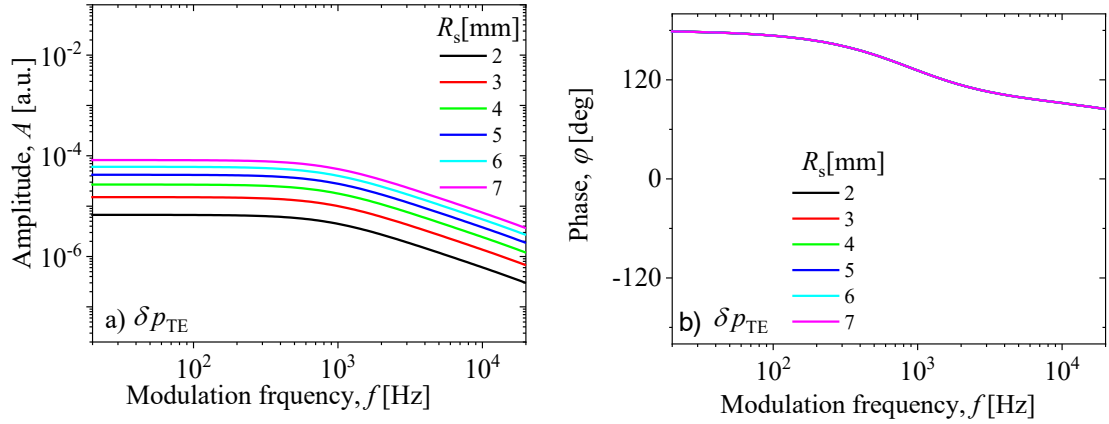

Figure 6. Thermoelastic component $\delta p_{TE}(f)$ a) amplitudes $A$ and b) phases $\varphi$ as a function of simple radius $R_s$, and modulation frequency, $f$.

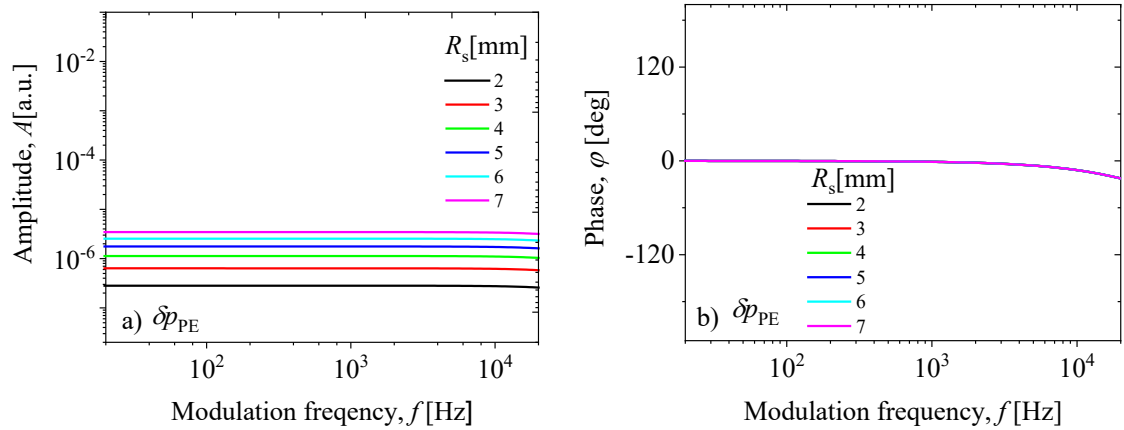

Figure 7. Plasmaelastic component $\delta p_{PE}(f)$ a) amplitudes $A$ and b) phases $\varphi$ as a function of simple radius $R_s$, and modulation frequency, $f$.

To fully understand what presented signal component amplitude and phase changes mean, it is necessary to draw all the components and the total signal together, because not only the values of the components are important but also their mutual relationship, which is shown in Figure 8. This figure shows the amplitudes and phases of total signals and their components for $R_s = 2$ mm (solid lines) and $R_s = 7$ mm (dashed lines). It is clear that the amplitude changes of the $\delta p_{PE}(f)$ component (blue) is smaller in absolute values compared to the changes in the $\delta p_{TE}(f)$ (green). This is the reason why $\delta p_{PE}(f)$ components do not affect the change in the total signal $\delta p_{total}(f)$ (black). Therefore, only the mutual relationship between $\delta p_{TD}(f)$ (red) and $\delta p_{TE}(f)$ (green) amplitudes drives the $\delta p_{total}(f)$ behaviour in frequency domain.

If $R_s = 2$ mm, the intersection of $\delta p_{TD}(f)$ and $\delta p_{TE}(f)$ is at $f \sim 2.4$ kHz, and therefore the $\delta p_{total}(f)$ (solid black) has a saddle shape at higher frequencies. When $R_s = 7$ mm, the intersection of $\delta p_{TD}(f)$ and $\delta p_{TE}(f)$ is at $f \sim 600$ Hz, and therefore $\delta p_{total}(f)$ (dashed black) is totally flattened (it has lost its saddle shape). It is clear that changes in the total signal for these and all other $R_s$ values used to make the presented base will be visible in the range $f > 200$ Hz. The same conclusion states for the distorted $S_{total}(f)$ signal, too. The analysis of the change of radius outside the range from 2 mm to 7 mm shows the loss of the influence of thermoelastic components on the total signal in the case of $R_s < 2$ mm, and their dominance in the total signal for $R_s > 7$ mm.

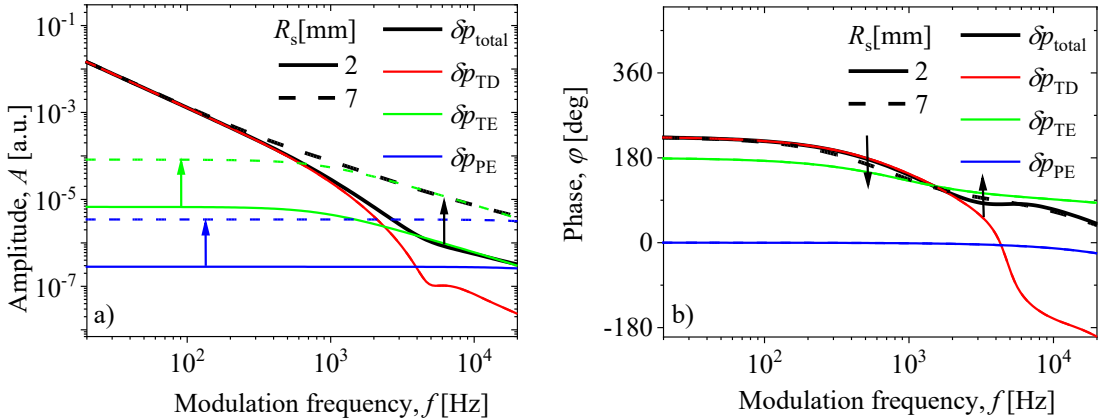

Figure 8. The tendency of changes in a) amplitudes $A$ and b) phases $\varphi$ of total photoacoustic signals $\delta p_{total}(f)$ (black lines) with its components: thermodiffusion $\delta p_{TD}(f)$ (red lines), thermoelastic $\delta p_{TE}(f)$ (green lines) and plasma elastic $\delta p_{PE}(f)$ (blue lines) for 400 μm.thick Si samples of radius $R_s$ = 2mm and $R_s$ = 7mm.

Presented theoretical analysis of the total signal behavior by changing the radius of the sample shows that it makes sense to use both undistorted and distorted total signals to recognize $R_s$ values because their amplitudes change significantly with changes in $R_s$ only at higher modulation frequencies. The low-frequency range can be used as the total signal self-normalization range, due to the absence of any $R_s$ influences on the total signal.

### 4.2. Network prediction results

The training results of NN1 neural networks trained by the base of distorted (Figure 3) and NN2 trained with the base of undistorted (Figure 4) signals are presented by the Mean Square Error (MSE) dependence of number of Epochs in Figure 9. During network training, bases are divided into a training base, a testing base, and a validation base, on the basis of which the generalized capabilities of networks are determined. In order to ensure the best performance of the networks, the criterion of interrupting the MSE deviation of individual curves is activated. NN1 network stopped its work in 3 epochs with the best validation performance of 0.014063 (Figure 9.a), while NN2 network stopped its work in 4 Epochs, with the best validation performance of 0.0075805 (Figure 9.b).

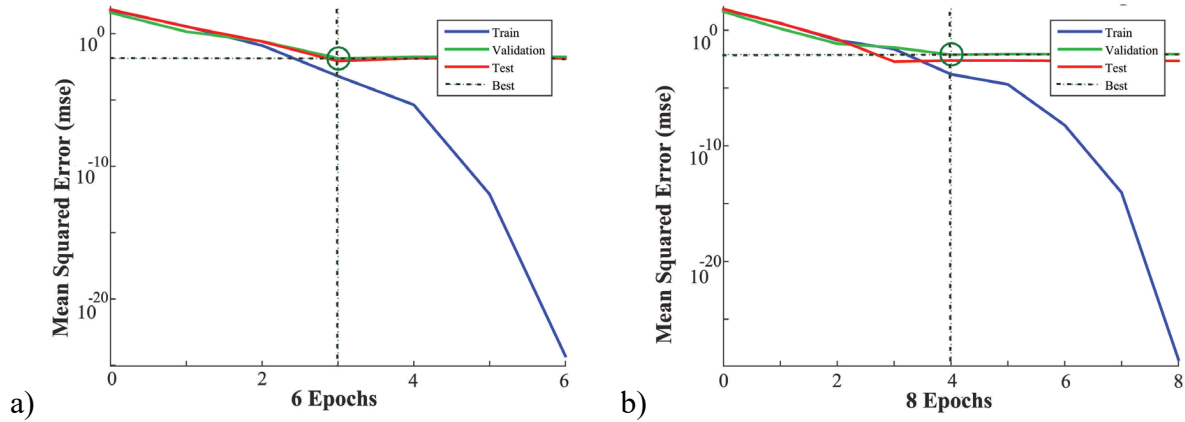

Figure 9. Network performance results: a) NN1 trained on distorted FA signals and b) NN2 trained on undistorted FA signals.

To evaluate the neural networks NN1 and NN2 ability to predict, we formed additional distorted and undistorted signal bases with values from 2.1 mm to 6.9 mm, formed using a step of 0.4 mm. These additional bases are used for network testing. In this way, we obtained 13 distorted +13 undistorted test signals evenly distributed throughout the range of $R_s$ changes. The results of the neural network test on these signals are shown in Table I. The obtained relative errors are at the experimental measurements level of accuracy, which shows that for the network analysis of certain properties of samples, large photoacoustic signal databases are not necessary.

For the sake of clarity, the results in Table I are presented in Figure 10. It can be clearly seen, based on the relative error percentage for NN1 and NN2, (Table I and Figure 10) that the prediction is good only in limited sample radius areas. A more accurate prediction of the NN1 (low % value of relative error) is for samples having $R_s \leq 3mm$ and samples with a radius larger than the microphone dimension ($R_s \geq 5mm$). A more precise prediction of the NN2 network ranges in a much wider range of radius values ($R_s \leq 4.5mm$), while for higher values of $R_s$ it gives poorer predictions. The largest (%) relative errors are made by both NN1 and NN2 networks for the radius of the sample in the range (4.0-5.5) mm, ie. about a value equal to the radius of the microphone $R_0 = 4.8$ mm. These results lead to the conclusion that both networks give the best predictions when the value of $R_s$ approaches the value of the microphone aperture radius $(r_0 = 1.5 \text{ mm})$. This confirms the justification of the approximations $R_s = r_0$ taken here with the experimental configuration of the open cell. In general, the results show that for experimental work it is best to take samples whose dimensions are smaller than the dimensions of the microphone ($R_0$).

Table I: The results of NN1 and NN2 neural networks predictions of sample radius given in the table with relative (%) errors. Asterisk (*) denotes the predictions for an experimentally measured PAS.

| $R_s$ [mm] | NN1 $R_s$ [mm] | rel % error, NN1 | NN2 $R_s$ [mm] | rel % error, NN2 |
|---|---|---|---|---|
| 2.1 | 2.1356 | 1.6933 | 2.1023 | 0.1086 |
| 2.5 | 2.4849 | 0.6048 | 2.5400 | 1.5996 |
| 2.9 | 2.9038 | 0.1319 | 2.8572 | 1.4745 |
| 3.3 | 3.2037 | 2.9183 | 2.8572 | 0.3332 |
| 3.7 | 3.8832 | 4.9527 | 3.7098 | 0.2650 |
| 4.1 | 4.1947 | 2.3089 | 4.1427 | 1.0416 |
| 4.5 | 4.7968 | 6.5967 | 4.5426 | 0.9470 |
| 4.9 | 4.8915 | 0.1740 | 4.5426 | 7.2936 |
| 5.3 | 5.3213 | 0.4014 | 5.1965 | 1.9530 |
| 5.7 | 5.6535 | 0.8155 | 5.6941 | 0.1028 |
| 6.1 | 6.0561 | 0.7190 | 6.1364 | 0.5964 |
| 6.5 | 6.4283 | 1.1034 | 6.1364 | 5.5942 |
| 6.9 | 6.8216 | 1.1358 | 6.8506 | 0.7156 |
| 4.54* | 4.7976 | 5.6749 | 4.6644 | 2.7398 |

To confirm the previous results, both networks presented the experimental results of distorted and undistorted signals of a sample that had a radius approximately equal to the microphone radius: $R_s \approx R_0$.

The results of network prediction in the case of the signal from Figure 11 obtained using the open photoacoustic cell experimental set-up are shown in Table I with an asterisk (*), and in Figure 10 with a dots (red for NN1, black for NN2). Based on the experimental results, it can be concluded that the prediction of the experimentally corrected photoacoustic signals by the NN2 network is more accurate. Therefore, it can be recommended that networks trained on undistorted signals (NN2) should be used in determining the $R_s$ of the tested sample only with the values smaller than the microphone radius $R_0$.

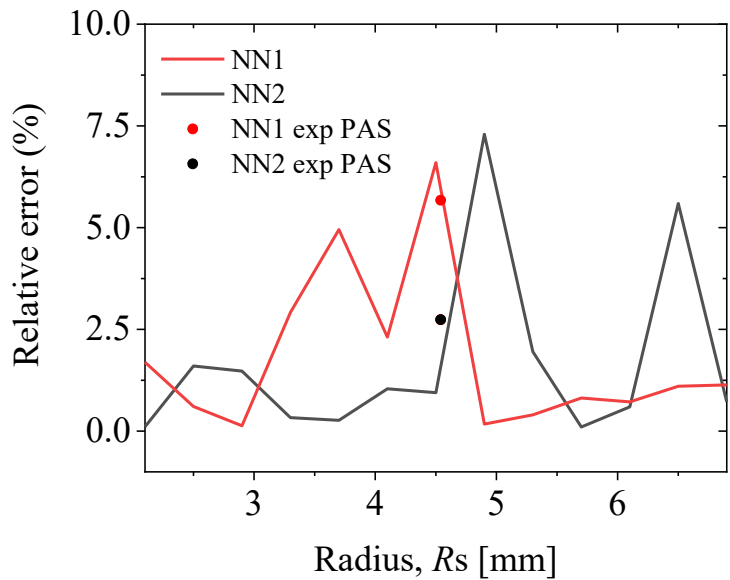

Figure 10. Dependence of relative% error prediction of NN1 (red line) and NN2 (black line) networks on the sample radius for the test of theoretical results and verification with experimental test with network NN1 (red dot), NN2 (black dot).

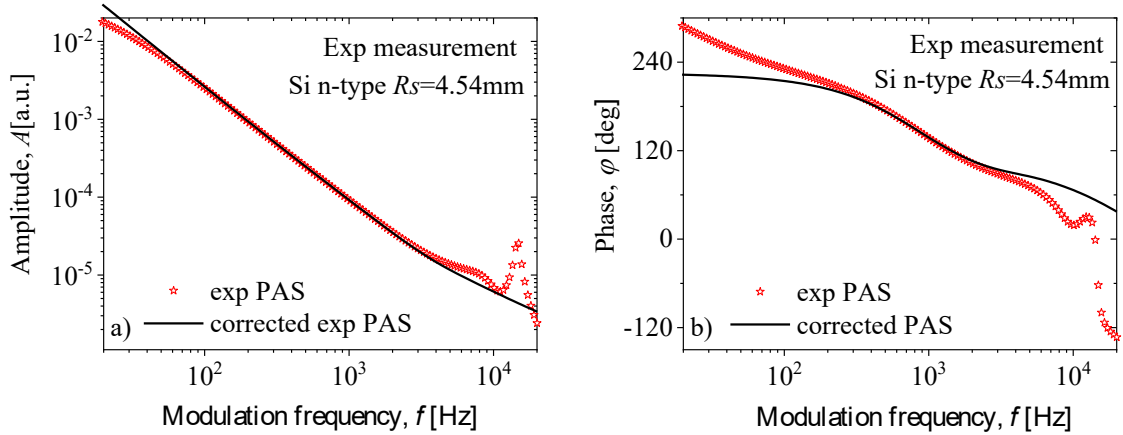

Figure 11. Experimental result of a) amplitude I b) phase of distorted (star) and undistorted (black line) photoacoustic signal presented to the NN2 and NN1, respectively, for a sample of 400 μm thick silicon n-type with a radius of $R_s$ = 4.54mm.

**Concusions**

We have shown that the radius of the sample $R_s$ is a very important parameter that can lead to changes in the shape of the amplitude and phase of the total photoacoustic signal and its elastic components (plasma elastic and thermoelastic) in the frequency domain from 20 Hz to 20 kHz. The analysis includes the total signals in respect to the influence of measuring instruments (distorted signals) and those generated directly by the illuminated sample (undistorted signals). It was found that the largest changes in the total signal due to $R_s$ changes occure in its thermoelastic component amplitude, bearing in mind that amplitude is proportional to $R_S^4$. It means that total signal amplitude changes are expected to find only at modulation frequencies $f$ > 100 Hz, while at lower frequencies the total signal amplitude remains the same, ie. it is not sensitive to $R_s$ changes at all. As it is expected, changes in the PE component amplitude are found to have no effect on total signal behaviour. It was also found that amplitudes of the signals that are distorted under the influence of measuring instruments have similar behavior due to $R_s$ changes. It was determined that the radius of the sample does not affect the change of the thermodiffusion component and that the largest changes in the total undistorted and distorted signals are due to changes in the relationship between the thermodiffusion and thermoelastic components, which are noticeable in the modulation frequency domain of $f$ > 100 Hz. The same relationship changes the phase of total signals in the entire frequency domain.

Numerical simulation created bases of distorted and undistorted signals depending on the sample radius in the range of changes from 2 mm to 7 mm with a change step of 0.2 mm, which were used for NN1 and NN2 network training, respectively. Through training, it has been shown that both networks achieve satisfactory accuracy after only a few epochs. This means that shallow networks can achieve acceptable performances even with databases created from a small amount of data. Undistorted and distorted signals, out of the base, having the same range of $R_s$ but with a different step of its changes are presented to such trained networks due to the test.

Predictions of both NN1 and NN2 networks proved to be the best for small values of $R_s$ in the range (2 - 3) mm, close to the dimensions of the microphone aperture ($r_0$ = 1.5 mm). This result indicates the justification of the theoretical approximation which assumes that $R_s = r_0$. In that case correction of the measured signal due to the instrumental influences is not necessary to obtain correct network prediction, which means that the experimental data analysis procedure can be simplified.

When $R_s$ values approaches the microphone radius $R_0$ region, the NN2 network trained with undistorted signals gives better predictions. Both networks give poorer predictions for samples whose radii are larger than $R_0$. This result leads to the conclusion that the samples used in open photoacoustic cell configuration, whose radii are not larger than the dimensions of the microphone should be used.

**Acknowledgments**
We are thankful for the financial support of this research by the Ministry of Education, Science and Technology development of the Republic of Serbia, contract number 451-03-09/2021-14/200017.

**Authors declarations** Conflict of Interest The authors have no conflicts to disclose.

**Data Availability** The data that support the findings of this study are available from the corresponding author upon reasonable request.

## Appendix I. Periodic temperature component in Si sample

Following the geometry of the problem presented in Figure 1, we are enabled to use the 1D diffusion equation to describe the temperature changes within the investigated Si sample along the heat propagation axis (z-axis) [8,32,33]:

$$\frac{d^2 T_s(z)}{dz^2} - \sigma^2 T_s(z) = -\frac{1}{k}\left[\frac{\varepsilon - \varepsilon_g}{\varepsilon}\beta I_0 \exp(-\beta z) - \frac{\varepsilon_g}{\tau}\delta n_p(z)\right]. \tag{A1.1}$$

The solution $T_s(z)$ to the Eq.(3) can be given as a sum of thermalization $T_{\text{therm}}(z)$, bulk $T_{\text{br}}(x)$ and surface $T_{\text{sr}}(x)$ recombination components [8,32]

$$T_s(z) = T_{\text{therm}}(z) + T_{\text{br}}(z) + T_{\text{sr}}(z), \tag{A1.2}$$

defined as

$$T_{\text{therm}}(z) = \frac{I_0}{k}\frac{\varepsilon-\varepsilon_g}{\varepsilon}\frac{\beta}{\beta^2-\sigma^2}\left[b\frac{e^{\sigma(z-l)}+e^{-\sigma(z-l)}-e^{-\beta l}\left(e^{\sigma z}+e^{-\sigma z}\right)}{e^{\sigma l}-e^{-\sigma l}}-e^{-\beta z}\right], \tag{A1.3}$$

$$T_{\text{br}}(z) = \frac{\varepsilon_g B_1}{\tau k \sigma^2}\left\{\frac{B_2 e^{\sigma z}+B_3 e^{-\sigma z}}{e^{\sigma l}-e^{-\sigma l}}-\frac{1}{c^2-1}\left[\frac{\delta n_p(z)}{B_1}+\frac{b^2-c^2}{b^2-1}e^{-\beta z}\right]\right\}, \tag{A1.4}$$

$$T_{\text{sr}}(z) = \frac{2\varepsilon_g}{k\sigma}\frac{s_1 \delta n_p(0)\cosh[\sigma(z-l)]+s_2 \delta n_p(l)\cosh(\sigma z)}{e^{\sigma l}-e^{-\sigma l}}, \tag{A1.5}$$

where constants are

$$b = \frac{\beta}{\sigma}, \quad c = \frac{1}{L\sigma},$$

$$B_1 = \frac{\beta I_0}{\varepsilon D_p(\beta^2 - L^{-2})}$$

$$B_2 = B_4 e^{-\sigma l} + B_5,$$

$$B_3 = B_4 e^{\sigma l} + B_5,$$

and

$$B_4 = -c\frac{\frac{1}{B_1}[\delta n_p(l) - \delta n_p(0)\cosh(l/L)] - \cosh(l/L) + e^{-\beta l}}{\sinh(l/L)\cdot(c^2-1)} - \frac{b}{b^2-1},$$

$$B_5 = c\frac{\frac{1}{B_1}\left[\delta n_p(l)\cosh(l/L) - \delta n_p(0)\right] - 1 + e^{-\beta l}\cosh(l/L)}{\sinh(l/L)\cdot(c^2-1)} + \frac{be^{-\beta l}}{b^2-1}.$$

Here, $\sigma = (1+i)/\mu$ being the complex heat diffusion coefficient, $\mu = \sqrt{D/\pi f}$ the thermal diffusion length of the semiconductor, $\beta$ the semiconductor absorption coefficient, and $D$ the thermal diffusivity. Also, $L = L_p/(1+i2\pi f\tau_p)^{1/2}$ is the complex excess carriers diffusion length, $L_p = \sqrt{D_p\tau_p}$ the minority excess carrier diffusion length, $i$ the complex unit, $\delta n_p(z)$ the photogenerated minority carriers (holes) dynamic density component, $D_p$ the holes diffusion coefficient, $\tau_p$ the holes lifetime, and $s_1$ and $s_2$ are the surface recombination velocities at front and rear sample surfaces, respectively.

Total temperature and its components at the rear (nonilluminated) side of the sample ($z=l, e^{-\beta l} \to 0$) can be written as [8,32]:

$$T_s(l) = T_{\text{therm}}(l) + T_{\text{br}}(l) + T_{\text{sr}}(l), \tag{A1.6}$$

where

$$T_{\text{therm}}(l) = \frac{I_0}{k\sigma}\frac{\varepsilon - \varepsilon_g}{\varepsilon}\frac{\beta^2}{\beta^2 - \sigma^2}\frac{1}{\sinh(\sigma l)}, \tag{A1.7}$$

$$T_{\text{br}}(z) = \frac{\varepsilon_g}{\tau k\sigma^2}\frac{\beta I_0}{\varepsilon D_p(\beta^2 - L^{-2})}\left\{\frac{B_2 e^{\sigma l} + B_3 e^{-\sigma l}}{2\sinh(\sigma l)} - \frac{1}{c^2-1}\frac{\delta n_p(z)}{B_1}\right\}, \tag{A1.8}$$

$$T_{\text{sr}}(l) = \frac{\varepsilon_g}{k\sigma}\frac{s_1\delta n_p(0) + s_2\delta n_p(l)\cosh(\sigma l)}{\sinh(\sigma l)}, \tag{A1.9}$$

**Appendix II. Periodic carrier densities in Si sample**

Considering that the used Si membranes (Figure 1) are the low-level injection *n*-type semiconductors (minority carriers are used to explain carrier dynamics), the dynamic part of the 1D diffusion equation (important for sound wave generation) that explains carrier transport in membranes can be written in the form [8,32,33]:

$$\frac{d^2\delta n_p(z)}{dz^2} - \frac{\delta n_p(z)}{L^2} = -\frac{\beta I_0}{\varepsilon D_p}e^{-\beta z}, \tag{A2.1}$$

where $L = L_p / (1 + i2\pi f \tau_p)^{1/2}$ is the complex excess carriers diffusion length, $L_p = \sqrt{D_p \tau_p}$ is the excess carrier diffusion length, $i$ is the complex unit, $\delta n_p(z)$ is the photogenerated minority (holes) dynamic density component, $D_p$ is the holes diffusion coefficient, and $\tau_p$ is their lifetime. The solution $\delta n_p(z, f)$ of the Eq.(17) is given by [8,32,33]:

$$\delta n_p(z) = A_+ e^{z/L} + A_- e^{-z/L} - A e^{-\beta z}, \qquad (A2.2)$$

where $A = I_0 / (\varepsilon D_p \beta)$, and the integration constants $A_\pm$ are defined by [8,32,33]:

$$A_\pm = \frac{A}{v_D} \frac{v_\beta (v_D \pm s_2) e^{\mp l_2/L} - v_D (v_\beta - s_2) e^{-\beta l_2}}{(v_D + s_2) e^{l_2/L} - (v_D - s_2) e^{-l_2/L}}, \qquad (A2.3)$$

depending strongly on the relative ratio of the characteristic diffusion speeds $v_D = D_p / L$ and $v_\beta = \beta D_p$.